\documentstyle[12pt,aaspp4]{article}

\begin{document}

\slugcomment{Accepted for publication in The Astrophysical Journal}

\title{Multi-Epoch Multiwavelength Spectra and Models\\ for Blazar 3C~279}

\author{R.~C.~Hartman\altaffilmark{1,2},
           M.~B\"ottcher\altaffilmark{3,44},
	   G.~Aldering\altaffilmark{4},
           H.~Aller\altaffilmark{5},
           M.~Aller\altaffilmark{5},
        D.~E.~Backman\altaffilmark{6},
	T.~J.~Balonek\altaffilmark{7},
        D.~L.~Bertsch\altaffilmark{1},
	S.~D.~Bloom\altaffilmark{8},
	   H.~Bock\altaffilmark{9},
	   P.~Boltwood\altaffilmark{10},
        M.~T.~Carini\altaffilmark{42},
	   W.~Collmar\altaffilmark{11},
	   G.~De~Francesco\altaffilmark{12},
        E.~C.~Ferrara\altaffilmark{31},
	   W.~Freudling\altaffilmark{13},
        W.~K.~Gear\altaffilmark{43},
        P.~B.~Hall\altaffilmark{16},
           J.~Heidt\altaffilmark{9},
           P.~Hughes\altaffilmark{5},
        S.~D.~Hunter\altaffilmark{1},
	   S.~Jogee\altaffilmark{17},
        W.~N.~Johnson\altaffilmark{18},
	   G.~Kanbach\altaffilmark{11},
	   S.~Katajainen\altaffilmark{19},
	   M.~Kidger\altaffilmark{20},
           T.~Kii\altaffilmark{21},
	   M.~Koskimies\altaffilmark{19},
           A.~Kraus\altaffilmark{22},
           H.~Kubo\altaffilmark{23},
           O.~Kurtanidze\altaffilmark{35},
	   L.~Lanteri\altaffilmark{12},
           A.~Lawson\altaffilmark{25},
        Y.~C.~Lin\altaffilmark{26},
           U.~Lisenfeld\altaffilmark{27},
	   G.~Madejski\altaffilmark{29},
           F.~Makino\altaffilmark{21},
	   L.~Maraschi\altaffilmark{15},
        A.~P.~Marscher\altaffilmark{30},
        J.~P.~McFarland\altaffilmark{31},
           I.~McHardy\altaffilmark{25},
        H.~R.~Miller\altaffilmark{31},
           M.~Nikolashvili\altaffilmark{24},
	   K.~Nilsson\altaffilmark{19},
        J.~C.~Noble\altaffilmark{30},
	   G.~Nucciarelli\altaffilmark{33},
	   L.~Ostorero\altaffilmark{12},
           E.~Pian\altaffilmark{34},
	   T.~Pursimo\altaffilmark{19},
        C.~M.~Raiteri\altaffilmark{12},
           W.~Reich\altaffilmark{22},
	   R.~Rekola\altaffilmark{19},
        G.~M.~Richter\altaffilmark{35},
        E.~I.~Robson\altaffilmark{36},
	   A.~Sadun\altaffilmark{37},
	   T.~Savolainen\altaffilmark{19},
           A.~Sillanp\"a\"a\altaffilmark{19},
           A.~Smale\altaffilmark{28},
	   G.~Sobrito\altaffilmark{12},
	   P.~Sreekumar\altaffilmark{38},
        J.~A.~Stevens\altaffilmark{14},
        L.~O.~Takalo\altaffilmark{19},
	   F.~Tavecchio\altaffilmark{15},
           H.~Ter\"asranta\altaffilmark{39},
        D.~J.~Thompson\altaffilmark{1},
           M.~Tornikoski\altaffilmark{39},
           G.~Tosti\altaffilmark{33},
           H.~Ungerechts\altaffilmark{27},
        C.~M.~Urry\altaffilmark{40},
           E.~Valtaoja\altaffilmark{19,32},
           M.~Villata\altaffilmark{12}
        S.~J.~Wagner\altaffilmark{9},
	A.~E.~Wehrle\altaffilmark{41},
        J.~W.~Wilson\altaffilmark{31}}

\altaffiltext{1}{Code 661, NASA/GSFC, Greenbelt, MD 20771}
\altaffiltext{2}{rch@egret.gsfc.nasa.gov}
\altaffiltext{3}{Department of Space Physics and Astronomy, Rice
				     University, Houston, TX 77005-1892}
\altaffiltext{4}{Lawrence Berkeley National Laboratory, Mail Stop
                           50-232, 1 Cyclotron Road, Berkeley, CA 94720}
\altaffiltext{5}{Astronomy Department, University of Michigan, Ann
                                                        Arbor, MI 48109}
\altaffiltext{6}{Physics and Astronomy Department, Franklin and Marshall 
		       College, P.O. Box 3003, Lancaster, PA 17604-3003}
\altaffiltext{7}{Department of Physics and Astronomy, Colgate University, 
                                  13 Oak Drive, Hamilton, NY 13346-1398}
\altaffiltext{8}{Hampden-Sydney College, Hampden-Sydney, VA}
\altaffiltext{9}{Landessternwarte K\"onigstuhl, 69117 Heidelberg, Germany}
\altaffiltext{10}{1655 Main St. Stittsville, Ontario K2S 1N6, Canada}
\altaffiltext{11}{Max-Planck-Institut f\"ur Extraterrestrische Physik,
                                  P.O. Box 1603, 85740 Garching, Germany}
\altaffiltext{12}{Osservatorio Astronomico di Torino, Strada
                           Osservatorio 20, I-10025 Pino Torinese, Italy}
\altaffiltext{13}{European Southern Observatory and Space Telescope - 
     European Coordinating Facility, Karl-Schwarzschild-Strasse 2, 85748
                                         Garching bei M\"unchen, Germany}
\altaffiltext{14}{Mullard Space Science Laboratory, University College 
                 London, Holmburg, St. Mary, Dorking, Surrey RH5 6NT UK}
\altaffiltext{15}{Osservatorio Astronomico di Brera, Via Brera 28, 
                                                I-20121 Milan, Italy}
\altaffiltext{16}{Princeton University Observatory and Pontificia 
    Universidad Cat\'{o}lica de Chile, Departamento de Astronom\'{\i}a y 
    Astrof\'{\i}sica, Facultad de F\'{\i}sica, Casilla 306, Santiago 22, 
    Chile}
\altaffiltext{17}{Division of Physics, Mathematics, and Astronomy, 
       MS 105-24, California Institute of Technology, Pasadena, CA 91125}
\altaffiltext{18}{Code 7651,
                    Naval Research Laboratory, Washington, DC 20375-5352}
\altaffiltext{19}{Tuorla Observatory, V\"ais\"al\"antie 20,
                                            FIN-21500 Piikki\"o, Finland}
\altaffiltext{20}{Instituto de Astrofísica de Canarias, Calle Vía Láctea, 
				      E-38200 La Laguna, Tenerife, Spain}
\altaffiltext{21}{Institute of Space and Astronautical Science, 
                  3-1-1 Yoshinodai, Sagamihara, Kanagawa 229-8510, Japan}
\altaffiltext{22}{Max-Planck-Institut f\"ur Radioastronomie, Auf dem
                                         H\"ugel 69, 53121 Bonn, Germany}
\altaffiltext{23}{Department of Physics, Faculty of Science,
                       Kyoto University, Kyoto 606-8502, Japan}
\altaffiltext{24}{Abastumani Observatory, 383762 Abastumani,
						     Republic of Georgia}
\altaffiltext{25}{Department of Physics and Astronomy, University of
                                                         Southampton, UK}
\altaffiltext{26}{W. W. Hansen Experimental Physics Laboratory, Stanford
                                          University, Stanford, CA 94305}
\altaffiltext{27}{IRAM, Avenida Divina Pastora 7, N.C., 18012 Granada,
                                                                   Spain}
\altaffiltext{28}{Code 662, NASA Goddard Space Flight Center, 
						     Greenbelt, MD 20771}
\altaffiltext{29}{Stanford Linear Accelerator Center, GLAST group,
                               2575 Sand Hill Road, Menlo Park, CA 94025}
\altaffiltext{30}{Institute for Astrophysical Research, Boston
                     University, 725 Commonwealth Ave., Boston, MA 02215}
\altaffiltext{31}{Department of Physics and Astronomy, Georgia State
                                           University, Atlanta, GA 30303}
\altaffiltext{32}{Department of Physics, University of Turku, Turku, Finland}
\altaffiltext{33}{Osservatorio Astronomico di Perugia, Via Bonfigli,
                                                    06123 Perugia, Italy}
\altaffiltext{34}{Astronomical Observatory of Trieste, 
                             Via G.B. Tiepolo 11, I-34131 Trieste, Italy}
\altaffiltext{35}{Astrophysikalisches Institut Potsdam, An der Sternwarte
                                              16, 14482 Potsdam, Germany}
\altaffiltext{36}{Centre for Astrophysics, University of Central
      Lancashire, Preston PR1 2HE; Joint Astronomy Centre, 660 North
                                       A`ohoku Place, Hilo, Hawaii 96720}
\altaffiltext{37}{University of Colorado at Denver, Department of Physics, 
                  Campus Box 157, P.O. Box 173364, Denver, CO 80217-3364}
\altaffiltext{38}{ISRO Satellite Center, Bangalore, India}
\altaffiltext{39}{Mets\"ahovi Radio Observatory, Helsinki University of
                                  Technology, 02540 Kylm\"al\"a, Finland}
\altaffiltext{40}{Space Telescope Science Institute, 3700 San Martin
                                              Drive, Baltimore, MD 21218}
\altaffiltext{41}{Jet Propulsion Laboratory, California Institute of
                    Technology, 4800 Oak Grove Drive, Pasadena, CA 91109}
\altaffiltext{42}{Dept. of Physics and Astronomy, Western Kentucky
                      University, 1 Big Red Way, Bowling Green, KY 42104}
\altaffiltext{43}{Department of Physics and Astronomy, Cardiff University, 
                                      P.O. Box 913, Cardiff CF2 3YB, UK}
\altaffiltext{44}{Chandra Fellow}

\clearpage

\begin{abstract}

Of the blazars detected by EGRET in GeV $\gamma$ rays,
3C~279 is not only the best-observed by EGRET, but also one of the
best-monitored at lower frequencies.  We have assembled eleven spectra,
from GHz radio through GeV $\gamma$ rays, from the time intervals of
EGRET observations.
Although some of the data have appeared in previous publications, most 
are new, including data taken during the high states in early 1999 and
early 2000.  All of the spectra show substantial $\gamma$-ray 
contribution to the total luminosity of the object; in a high state, the
$\gamma$-ray luminosity dominates over that at all other frequencies by
a factor of more than 10.  There is no clear pattern of time correlation;
different bands do not always rise and fall together, even in the
optical, X-ray, and $\gamma$-ray bands.

The spectra are modeled using a leptonic jet, with combined synchrotron
self-Compton + external Compton $\gamma$-ray production.  
Spectral variability of 3C~279
is consistent with variations of the bulk Lorentz factor of the jet,
accompanied by changes in the spectral shape of the electron
distribution.  Our modeling results are consistent with the UV
spectrum of 3C~279 being dominated by accretion disk radiation
during times of low $\gamma$-ray intensity.

\end{abstract}

\keywords{quasars: individual (3C~279)}

\clearpage

\section{Introduction}

Partly because of its bright $\gamma$-ray flare shortly after the launch 
of the Compton Gamma Ray Observatory (Hartman et al. 1992; Kniffen et al. 
1993; Hartman et al. 1996), 3C~279 has received a great deal of coverage
during the following nine years, at many frequencies in addition to the
$\gamma$-ray range.  In addition to much routine monitoring, it has also
been the subject of six intensive multiwavelength campaigns.
It has been detected by the EGRET high-energy instrument on CGRO every 
time it has been observed, even (at rather low significance) during the
very low quiescent state of winter 1992-1993.  It has varied
in $\gamma$-ray flux by roughly two orders of magnitude (Maraschi et al.
1994; Wehrle et al. 1998), and has displayed factor-of-two variation on 
timescales as small as 8 hours.

Its detection by EGRET during the very low state of 1992 December -- 1993 
January (Maraschi et al. 1994) is undoubtedly due to the fact that 
3C~279 is one of the closest ($z = 0.538$) of the EGRET-detected 
flat-spectrum radio-loud quasars (FSRQ), also described recently as 
quasar-hosted blazars (QHB; Kubo et al. 1998).

In this paper we have collected multiwavelength data simultaneous or
(for radio) quasi-simultaneous with gamma-ray observations in order to
establish a detailed model picture for this prototype gamma-ray loud 
AGN, and to constrain the dynamics and physics of relativistic jets.

\section{Observations}

This paper concentrates on the time intervals during which the CGRO
instruments COMPTEL and EGRET were observing 3C~279.  In some cases,
adjacent CGRO viewing periods have been combined since the $\gamma$-ray
flux did not vary significantly.  The intervals used are defined
in Table 1.

Because the radio fluxes do not vary as rapidly as those at higher
frequencies, we have utilized a few radio observations that are a 
few days outside the time intervals listed in Table 1.  For frequencies 
above 70 GHZ all of the observations used were strictly within the 
listed time intervals.

Data points have been taken from three earlier papers: Maraschi et al.
(1994) for the 1992 December -- 1993 January (P2) spectrum, Hartman et al.
(1996) for the 1991 June (P1) spectrum, and Wehrle et al. (1998) for the 
two spectra (P5a and P5b) in 1996 January -- February.  For those four 
spectra the EGRET and COMPTEL $\gamma$-ray data have been reprocessed 
and reanalyzed for use here. 
 
Data were also obtained from the references and observations described
below.

\subsection{Radio}

The flux data at 4.8, 8.0, and 14.5 GHz were obtained with the University 
of Michigan 26 m paraboloid using rotating-horn polarimeters.  For each 
day's source observation, a standard noise signal was injected to 
determine the radiometer gain, position scans were made in right ascension  
and declination to provide the best source position, and a series of 2.5 
minute on-off (4.8 GHz) or on-on (14.5 and 8.0 GHz) measurements were made. 
The typical duration of a full observation of 3C~279 is 30--40 minutes per 
day. The  primary flux standard used is Cassiopeia~A (3C~461) (allowing 
for the measured decay rate of this source) on the scale of Baars et al. 
(1977).  Secondary flux standards (3C~274 and 3C~286) were observed at 
1.5--2 hour intervals near 3C~279 observations to monitor temporal changes 
in the gain of the telescope.  Details of these calibration and reduction
procedures can be found in Aller et al. (1985).

Observations made during P2, P3a, P3b, and P4 at Effelsberg have been
published in Reich et al. (1993, 1998).  The flux tables are available
at http://cdsweb.u-strasbg.fr/Abstract.html .  Additional measurements 
at Effelsberg were obtained within a long-term program of
flux density monitoring of selected AGN (Peng et al. 2000).

Radio observations at 22 and 37 GHz were carried out at the Mets\"ahovi 
Radio Observatory as a part of a long term monitoring program.  The 
1999 and 2000 data have not been published previously; earlier data were 
presented by Maraschi et al. (1994), Hartman et al. (1996), 
Ter\"asranta et al. (1998; the observations and data
reduction are described in full there), and Wehrle et al. (1998).

\subsection{Millimeter and Submillimeter}

3 mm and 1.3 mm observations were made with the 15 m
Swedish-ESO Submillimetre Telescope (SEST) at the European
Southern Observatory site on Cerro La Silla in Chile.  Until June 1995 
the 3 mm observations were made using a dual-polarization Schottky 
receiver; after that the measurements were made with an SIS receiver.
As a backend, a wide-band (1 GHz) acousto-optic spectrometer (AOS) was
used.  For the 1.3 mm observations, since 1991 measurements have been
mainly obtained with a single channel bolometer. The bolometer has
a bandwidth of about 50 GHz, centered at 236 GHz.
All the flux density measurements were made in a dual beamswitching
mode, and the flux densities were calibrated against planets.
Details about the SEST continuum observations as well as the full data
base of the project through 1994 June can be found in Tornikoski et
al. (1996).

Flux densities at 3 mm, 2 mm, and 1.3 mm wavelength were determined
from observations at the IRAM 30m telescope applying procedures,
calibration, and data reduction similar to those described by Steppe
et al. (1988) and Reuter et al. (1997).  During regular observations 
made to update the pointing model of the
telescope or during VLBI campaigns, we make cross-scans over the
source in azimuth and elevation, using SIS 
(superconductor-insulator-superconductor) 
receivers at all three wavelengths.  The data are
calibrated to the scale of corrected antenna temperature, $T_A^*$, by
observing loads at ambient and cold temperature, as in the
conventional "chopper-wheel" calibration for millimeter wavelength
observations.  The peak intensity is determined by fitting a Gaussian
to the scans in both directions, and corrected for the observed
position offsets.  Another correction accounts for the
elevation-dependence of the telescope gain for a point source.
Finally the conversion factor from $T_A^*$ to flux density is measured
using the same observing mode on Mars or Uranus, or on selected
Galactic sources, i.e., K3-50A, NGC 7027, NGC 7538, or W3OH, which
are used as secondary standards.

Two 3C~279 fluxes (86 GHz in P5a; 90 GHz in P8) were used from the 
BIMA (Welch et al. 1996) on-line calibration archive 
(http://bima2.astro.uiuc.edu/calibrators/).

3C~279  was observed  with the  OVRO millimeter-wave
interferometer (Padin et al. 1991) on 1994 December 8, 1999 January 8, 
1999 January 11, 1999 January 20, and 1999 February 16.
These observations were part of normal calibration  measurements
taken during science tracks in order to correct for temporal phase
variation and gain variation  across the spectrometer band.
Continuum data was recorded with  an analog continuum correlator
using a setup which provides a bandwidth of 1  GHz  on each sideband.
The  upper sideband was centered on frequencies of  114.68 and 228.17 GHz.
Stringent requirements on coherence, system temperatures, and elevation were
used to select the best data,  and the  flux calibration of 3C~279 was
carried out with the  Owens Valley millimeter array software  (Scoville
et al. 1993). The tracks did not include  planet observations; therefore
the quasar 3C~273 was used to determine the absolute flux scale.

JCMT data for P1, P2, and P5b were taken from Hartman et al. (1996),
Maraschi et al. (1994), and Wehrle et al. (1998), respectively.  
For P3a, P3b, and P4, the UKT14 bolometer (Duncan et al. 1990) was
used on the JCMT (Matthews 1991) to observe 3C~279
at 2 mm, 1.3 mm, 1.1 mm, 0.8 mm, and 0.45 mm.  The data analysis was
similar to standard photometric studies at optical wavelengths;
details of the techniques used can be found in Robson et al. (1993),
who present the results of very similar observations.  The planets 
Uranus and Mars were used as primary flux density calibrators
whenever possible, and strong unresolved sources were observed regularly
to correct the pointing and focus of the antenna. To correct for 
atmospheric extinction, a plane-parallel atmosphere was assumed, and
optical depths were calculated directly from observations of the 
calibrators, which were observed on roughly an hourly basis throughout 
each session.

Submillimeter data for P6b and P8 were obtained using the
Submillimetre Common-User Bolometer Array, SCUBA (Holland et al. 1999)
on the JCMT.
The observations are from two programs: (1) a dedicated blazar monitoring
program with SCUBA in photometry mode. The data reduction method is almost
identical to that used for UKT14 although for observations made with the
arrays, off-source bolometers can be used to correct for sky noise; and (2)
automated reduction of archival 850 $\mu$m pointing observations (Robson,
Stevens \& Jenness, in prep.). The data reduction is similar to
conventional methodology.  As described in Jenness et al. (in prep.),
it uses the new JCMT SCUBA pipeline with automatic atmospheric
extinction correction.  This comes from an extensive database of extinction
estimates to correct for the atmosphere, and reduces reliable observations
only, using a standard flux conversion factor determined from observations
of Uranus and CRL618. This factor was found to be stable to within 10 per
cent over a 3 year period.

\subsection{Infrared}

The infrared observations during P1, P2, P5 (a and b) were taken from
Hartman et al. (1996), Maraschi et al. (1994), and Wehrle et al.(1998),
respectively.

The P5b (1996 January 31 and Februay 3) CTIO observations used the
1.5 m telescope and near-IR camera CIRIM in photometric
conditions.  The data were reduced using apphot routines in the
IRAF software package.  Flux densities were obtained by reference
to four UKIRT faint standard stars using normal CTIO extinction
coefficients and CIRIM color coefficients.

The UKIRT observation during P3a (1994 December 9)
used the UKT9 1--5 $\mu$m single-channel photometer under photometric 
sky conditions.  The flux was obtained by comparison with nearby 
standard stars, with uncertainties in the magnitudes of 0.03--0.05.

\subsection{Optical}

3C~279 was observed during P8 (1999 Jan 18 -- Feb 13) and P9 
(2000 Jan 29 -- Feb 19) at the Abastumani Astrophysical Observatory
(Republic of Georgia) using a Peltier-cooled ST-6 CCD camera
attached to the Newtonian focus of the 70 cm meniscus telescope (1/3).
The full frame field of view is 14.9x10.7 arcmin$^2$.
All observations are performed using combined filters of glasses which
match the standard B, V (Johnson) and $R_C, I_C$ (Cousins) bands well.
Because the scale of the CCD and the meniscus telescope resolution are
2.3x2.7~arcsec$^2$ per pixel and 1.5 arcsec respectively, the images
are undersampled; therefore the frames were slightly defocused to satisfy
the sampling theorem.  A full description of the Abastumani blazar
monitoring program is given in Kurtanidze \& Nikolashvili (1999).

P5a and P5b observations at the Boltwood Observatory, presented
previously in Wehrle at al. (1998),
were made with an 18 cm refractor, using Johnson-Cousins BVRI 
filters and a Thomson CSF TH7883 CCD camera.  Photometry was 
done differentially based on star 1 of Smith \& Balonek (1998).

A P5a observation at NURO (1996 January 12) used the 0.8 m
telescope and a Tektronix CCD camera with Johnson-Cousins
BVRI filters.  Photometric reference stars were Landolt equatorial
standards (Landolt 1992).  Atmospheric
extinction coefficients were determined for that night from
observations of standards.  This observation was presented 
previously in Wehrle et al. (1998).

Observations were taken with the 1.2 m telescope of Calar Alto 
Observatory, Spain, and with the 0.7 m telescope of the 
Landessternwarte Heidelberg.  Both telescopes are equipped with 
$LN_2$-cooled CCD cameras.  Observations in Heidelberg are carried 
out with a Johnson R band filter.  The Calar Alto observations were
carried out in Johnson R (in 2000) and R\"oser R (in earlier years). 
Standard de-biasing and flat-fielding was carried out before 
performing differential aperture photometry.  (Finding charts and
comparison sequences are available at
http://www.lsw.uni-heidelberg.de/projects/extragalactic/charts.html
for 3C~279, along with many other blazars.)

Observations were performed using Lowell Observatory's 42 inch
Hall telescope and the 24 inch telescope of the Mount Stromlo / 
Siding Spring Observatories.  Both telescopes are equipped with a 
direct CCD camera and an autoguider. The observations were made
through VRI filters.  Repeated exposures of 90 seconds were obtained
for the starfield containing 3C~279 and several comparison stars 
(Smith et al. 1985).  These comparison
stars were internally calibrated and are located on the same CCD
frame as 3C~279.  They were used as the reference standard stars in 
the data reduction process.  The observations were reduced following
Noble et al. (1997), using the method of Howell and Jacoby
(1986).  Each exposure is
processed through an aperture photometry routine which reduces the 
data as if it were produced by a multi-star photometer. Differential
magnitudes can then be computed for any pair of stars on the frame.
Thus, simultaneous observations of 3C~279, several comparison stars,
and the sky background permit removal of variations which may be
due to fluctuations in either atmospheric transparency or extinction.
The aperture photometry routine used for these observations is the
{\it phot} task in IRAF.

Observations were taken with the 2.5 m Nordic Optical Telescope (NOT) 
on La Palma, Canary Islands, Spain, using the ALFOSC intrument with 
a 2000x2000 CCD camera (0.189 arcsecond per pixel), and V and R filters.
Data reduction (including bias amd flat field corrections) were made
either with standard IRAF or MIDAS (J. Heidt) routines.
The data taken in 1999 and 2000 are previously unpublished.

Observations at the Tuorla Observatory used the 1.03 m telescope with a
ST-8 CCD camera and a V-filter (Katajainen et al. 2000).  Data 
reduction was done using IRAF (with bias and flatfield corrections).
Some portions of the Tuorla data have been previously published in
Wehrle et al. (1998) and Katajainen et al. (2000).

Observations at the Perugia Observatory were carried out with the
Automatic Imaging
Telescope (AIT). The AIT is based on an equatorially mounted 40~cm
f/5 Newtonian reflector, with a 15~cm f/15 refractor solidly
joined to it. A CCD camera and Johnson-Cousins $BVR_cI_c$ filters are
utilized for photometry (Tosti et al 1996).  The data were reduced
using aperture photometry with the procedure described in that
reference.

Observations at the Torino Observatory were done with the 1.05 m REOSC 
telescope. The equipment includes an EEV CCD camera ($1296\times 1152$ 
pixels, 0.467 arcsec per pixel) and standard (Johnson-Cousins) $BVRI$ 
filters. Frames are reduced by the Robin procedure locally developed 
(Lanteri 1999), which includes bias subtraction, flat fielding, and 
circular Gaussian fit after background subtraction.  The magnitude 
calibration was performed according to the photometric sequence by 
Raiteri et al. (1998). Magnitudes were converted to fluxes by using a
$B$-band Galactic extinction of 0.06 mag and following Rieke \& Lebofsky 
(1985) and Cardelli et al. (1989).  Data from 1994--1995 were published 
in Villata et al. (1997); data around P5a and P5b were partially shown 
in Wehrle et al. (1998) and Villata et al. (1998).

P9 observations were made with the 60~cm KVA telescope on La Palma,
Canary Islands, using a ST-8 CCD camera with BVR filters.  The data 
reduction was done using IRAF (with bias and flatfield corrections). 
These data are previously unpublished.

\subsection{UV}

IUE data for P1 and P2 appeared in Hartman et al. (1996) and Maraschi 
et al. (1994), respectively, but were reanalyzed by Pian (1999).  
The IUE data for P3b and P4 were presented in Koratkar et al. (1998)
and Pian et al. (1999), and those of P5a and P5b were presented in 
Wehrle et al. (1998).

\subsection{X-Ray}

ROSAT data for P2 and P5a were presented in
Maraschi et al. (1994) and Wehrle
et al. (1998), respectively, but were reanalyzed by Sambruna (1997).

ASCA data for P3b and P5a were presented in Wehrle et al (1998) and 
Pian et al. (1999), respectively; they were reanalyzed by Kubo (1997).  
ASCA data for P4 have appeared only in Kubo (1997).

3C~279 was observed by BeppoSAX during 5 pointings in P6a (1997
Jan 11--21), for a total exposure time of 150 ks.
The spectra taken by the BeppoSAX LECS (0.1--4 keV; Parmar et al. 1997) and
MECS (1.6--10 keV; Boella et al. 1997) instruments have been extracted from
the linearized event files using extraction radii of 8$^{\prime\prime}$ and
4$^{\prime\prime}$, respectively, and corrected for background
contamination using library files available at the BeppoSAX Science Data
Center.

The spectra measured by the BeppoSAX PDS instrument (13--300 keV; Frontera 
et al.  1997) have been accumulated from on-source exposures of the 
collimator units, and corrected using background spectra obtained during 
the off-source exposures. A further correction for the energy and 
temperature dependence of the pulse rise time, devised for weak sources, 
has been applied.

No emission variability has been detected among different pointings,
therefore the spectral signal has been coadded over all exposures to
maximize the S/N.  The average spectrum of the source exhibits significant
signal up to $\sim$50 keV and appears featureless; see Pian (in 
preparation) for a detailed presentation of these data.
			       
After rebinning the spectrum in intervals where the signal exceeds a
3$\sigma$ significance, we fitted it to a single absorbed power-law using
the XSPEC routines and response files available at the Science Data
Center.  The normalization between the MECS and the LECS
instruments has been treated as a free parameter, while that between the
MECS and PDS instruments was fixed.  The fit is satisfactory; 
the fitted neutral hydrogen column density is consistent with the
Galactic value (Elvis, Lockman, \& Wilkes 1989) and the L/M normalization
is consistent with the expected intercalibration between the two
instruments.  

Ginga fluxes for P1 were taken from Hartman et al. (1996).

RXTE data from P5a and P5b have been presented previously in Wehrle et
al. (1998).  Full details of those analyses are given in
Lawson et al. (1999).  Subsequent data
were analysed in almost exactly the same manner but, for completeness,
we summarise the observations and analysis briefly.

3C~279 was the target for a series of 36 RXTE monitoring observations
during 1999 January 2 -- February 16, for a total on-source time
of 67 ksec.  The X-ray data presented here were obtained using the
Proportional Counter Array (PCA) instrument in the ``Standard~2'' and 
``Good Xenon'' configurations, with time resolutions of 16 sec and 
$<1\mu$sec respectively.  Only PCUs 0, 1, and 2 were
reliably on throughout the observations, and we limit our analysis to
data from these detectors.

A further sequence of 28 monitoring observations was performed with
RXTE in 2000 February, using the same instrumental configurations, for
a total on-source time of 104 ksec.  For this sequence we utilized
data from PCUs 0 and 2.
 
Data analysis was performed using RXTE standard analysis
software, FTOOLS 5.0.  Background subtraction of the PCA data was
performed utilizing the ``L7-240'' models generated by the RXTE
PCA team.  The quality of the background subtraction was checked in
two ways: (i) by comparing the source and background spectra and light
curves at high energies (50--100 keV) where the source itself no longer
contributes detectable events; and (ii) by using the same models to
background-subtract the data obtained during slews to and from the
source.

\subsection{Gamma-Ray}

OSSE consists of four identical scintillation detectors which make
measurements in the 50~keV -- 10~MeV gamma ray energy band.  The 
detectors are actively shielded and the rectangular fields of view, 
$3.8\arcdeg \times 11.4\arcdeg$ full width at half-maximum, are 
defined by tungsten collimators.  OSSE utilizes
a sequence of two-minute source observations alternated with 
offset-pointed background observations to determine a source 
contribution above background.  Generally, equal times are spent on 
source and background measurements.  See Johnson et al.\ (1993) for 
a more detailed description of the instrument, its
performance and data analysis techniques.

The background-subtracted data from each observation period are 
summed to obtain a time-averaged count rate spectrum for that period.
Prior to fitting, systematic errors are included.  These systematics 
are computed from the uncertainties in the low energy calibration 
and  response of the detectors using both in-orbit and pre-launch 
calibration data.  The energy-dependent systematic errors are 
expressed as an uncertainty in the effective area in the
OSSE response.  These systematic errors were added in quadrature to 
the statistical errors prior to spectral fitting.  They are most 
important at the lowest energies ($\sim 3\%$ uncertainty in 
effective area at 50 keV, decreasing
to 0.3\% at 150 keV and above).
 
Simple spectral models are fit to the residual count spectra from 
each viewing period or to the sum of several
viewing periods using the OSSE data analysis system.  The fitting 
procedure consists of folding model photon spectra through the OSSE 
response matrix and adjusting the model parameters to minimize, 
through a $\chi^2$ test, the deviations between the model count 
spectrum and the observed count spectrum.  For observations with 
insignificant count data, upper limits to the flux are
derived by fitting a power law spectrum with a photon index of 2.0.

OSSE results for P5a and P5b were presented in Wehrle et al. (1998),
but were reanalyzed for this work.
For P3b, P6b, P8, and P9, the OSSE 50--150 keV fluxes or upper limits
were obtained from routine automatic processing.  For P2, P4, and P6a,
the standard analysis was performed on sums of several viewing 
periods.

The COMPTEL experiment, mounted parallel to EGRET on CGRO,
always observes simultaneously with EGRET, but is sensitive to
lower-energy $\gamma$-rays, from 0.75~MeV to 30~MeV.
For details about the COMPTEL instrument see Sch\"onfelder et al. (1993).
The COMPTEL data have been analyzed following the standard
COMPTEL procedures (see e.g. Collmar et al. 2000a and
references therein). The COMPTEL data of 1991 (P1) were previously
presented in Hermsen et al. (1993), Williams et al. (1995), and
Hartman et al. (1996); the data of 1993 (P3a, P3b)
in Collmar et al. (1996); the data of 1996 (P5a, P5b) in Collmar et 
al. (1997) and Wehrle et al. (1998); and those of 1999 (P8)
in Collmar et al. (2000b).

The EGRET instrument on CGRO is sensitive to $\gamma$-rays in the energy
range 30 to 30,000~MeV.  Its capabilities and calibration are described
in Thompson et al. (1993) and Esposito et al. (1999).  Point source data
are analyzed using likelihood techniques (Mattox et al. 1996).
The data from P1 have been previously presented in Hartman et al. (1996);
those from P2 in Maraschi et al. (1994); and those from P5a and P5b
in Wehrle et al. (1998).

For the present work, the data from COMPTEL, EGRET, and OSSE have
been completely reprocessed and reanalyzed; the results of the reanalysis 
are consistent with the earlier publications, although not identical.

\section{Gamma-Ray Light Curve and Multiwavelength Spectra}

In order to illustrate the general activity level of 3C~279 at the time
of each of the spectra shown below, the light curve of all EGRET 
observations of 3C~279 ($E > 100 MeV$) is 
shown in Figure 1.  The fluxes shown vary by a factor of about 56, which
is greater than the variation observed in any other frequency band
over this time span.  (Note that the $\gamma$-ray fluxes shown are from
integrations over at least one week.  However, the one-day flux for
4~Feb~1996 was greater than 1000 on the same scale, making the total 
range of variation well over a factor of 100.)  Dramatic variability 
on short time scales is a well-known characteristic of this object at
all frequencies (see, e.g., Webb et al. 1990), but the amplitude seems
to increase with frequency.

The eleven multiwavelength spectra assembled from the observations
described above are shown in Figure 2(a--k) with the model fits described
in the next section.  The CGRO observation during October 1991 is not
included in that figure, because 3C~279 was too close to the Sun for
observations to be made at lower frequencies.

In the Web-based version of this paper (http://etcetc), Tables P1, P2,
P3a, P3b, P4, P5a, P5b, P6a, P6b, P8, and P9
give the values of the data points plotted in Figure 2(a--k),
in a form suitable for computer download.

\section{Modeling}

The broadband spectra were modeled with the pair plasma jet simulation
code described in detail in B\"ottcher et al. (1997) and 
B\"ottcher \& Bloom (2000).  In this model, blobs of
ultrarelativistic pair plasma are moving outward from the central
accretion disk along a pre-existing straight cylindrical jet structure, 
with relativistic speed $\beta c$ and bulk Lorentz factor $\Gamma$ 
(see Figure 3).  In the following, primed quantities refer 
to the rest frame comoving with the relativistic plasma
blobs.  At the time of injection into the jet at height $z_i$ above the
accretion disk, the pair plasma is assumed to have an isotropic 
power-law energy distribution in the comoving
frame, with index $p$ and low- and high-energy cut-offs 
$\gamma_1 \le \gamma' \le \gamma_2$.  The blobs are spherical
in the comoving frame, with radius $R'_B$, which does not change along 
the jet.  A randomly oriented magnetic field $B' \lesssim B'_{ep}$ 
is present, where $B'_{ep}$ is the magnetic field corresponding
to equipartition with the energy density of pairs at the base of the
jet.  The observer is misaligned by an angle $\theta_{obs}$ with 
respect to the jet axis.

As the blob moves out, various radiation and cooling mechanisms are at
work: synchrotron radiation, synchrotron self-Compton radiation 
(SSC; Marscher \& Gear 1985, Maraschi et al. 1992, Bloom \& 
Marscher 1996), inverse-Compton scattering of external radiation from
the accretion disk, either entering the jet directly (ECD, for External
Compton scattering of direct Disk radiation; Dermer et al. 1992, 
Dermer \& Schlickeiser 1993) or after being rescattered by surrounding 
(broad-line region) material (ECC for External Compton
scattering of radiation from Clouds; Sikora et al. 1994, Blandford 
\& Levinson 1995, Dermer et al. 1997).

The contribution from Comptonization of infrared emission
from heated dust in the circumnuclear region can become
quite significant and even dominant over the ECC radiation
in the soft gamma-ray regime in the case of continuous
reinjection or reacceleration of relativistic pairs in the jet,
because the energy density of this radiation field remains
approximately constant over a much longer length scale
than the radiation field due to accretion disk radiation
reprocessed in the BLR (Blazejowski et al. 2000).
However, in the case of instantaneous
particle injection and subsequent cooling, considered in
the model applied in this paper, that contribution is
negligible since in this case the relativistic electrons
have transformed virtually all their kinetic energy into
radiation by the time the blob leaves the BLR.

The broad-line region (BLR) is represented by a spherical layer of 
material with uniform
density and total Thomson depth $\tau_{T,BLR}$, extending between the 
radii $r_{in, BLR} \le r \le r_{out, BLR}$ from the central black
hole.  The central accretion disk is radiating a standard
Shakura-Sunyaev (1973) disk
spectrum around a black hole of $1.5 \cdot 10^8$ solar masses.  
The total disk luminosity is assumed to be $10^{46}$~erg~s$^{-1}$.  
The full angle dependence of the disk radiation
field, according to its radial temperature structure, is taken 
into account.  Compton scattering of all radiation
fields is calculated using the full Klein-Nishina cross section.  The
SSC mechanism is calculated to arbitrarily high scattering order.

After the initial injection of ultrarelativistic pairs, all of the
mechanisms mentioned above are taken into account in order to 
follow the evolution of the electron/positron distribution
functions and the time-dependent broadband emission as the blob moves
out along the jet.  $\gamma\gamma$ absorption intrinsic to the 
source and the re-injection of pairs due to $\gamma\gamma$ pair 
production is taken into account completely self-consistently 
using the exact analytic solution of B\"ottcher \& Schlickeiser 
(1997) for the pair injection spectrum due to $\gamma\gamma$ 
absorption.

Given the necessary integration times of at least several days for 
EGRET to obtain spectral information (longer than the cooling time 
of pairs in the jet), we use the time-integrated
spectrum of our simulated relativistic blobs, and re-convert the
resulting fluence into a flux by dividing it by an average 
repetition time of blob ejection events.  This repetition time
determines the overall normalization of our model spectra, which is
constrained by the requirement that subsequent blobs are not 
allowed to overlap.

The parameters that were fixed for modeling all eleven spectra are 
shown in Table 2.  These are thought to be fairly well constrained 
by other observations, or do not appear to affect significantly the 
values obtained for the parameters obtained from the modeling, 
which are shown in Table 3.

The model used here addresses only emission from the innermost 
portion of the jet.  In that region, the centimeter-wavelength 
radio emission is strongly synchrotron self-absorbed.  The 
centimeter fluxes seen in the spectra of 
Figure 2 are believed to come from much farther out along the jet, 
i.e., one to several parsecs from the central engine.  That 
parsec-scale radio emission has been modeled by, for example, Hughes 
et al. (1991) and Carrara et al. (1993).  With those models, much 
larger angles of the jet to the line of sight (30$^\circ$ to 
40$^\circ$) have been found to be appropriate.  
The inner jet must be directed at smaller
angle to the line of sight to be consistent with the
observed superluminal motion.  This change can be 
attributed to bending of the jet, which is seen as a common feature 
in VLBI maps of blazars.

In general, the millimeter radio emission in blazars seems to be more 
closely connected to the gamma-ray variations, being either cospatially
or slightly farther out along the jet (Reich et al. 1993; Valtaoja
\& Ter\"asranta 1996).  Combining 
VLBI and total flux density variation data, L\"ahteenm\"aki \&
Valtaoja (1999) estimated that the jet Lorentz factor is about 9 
and the viewing angle 2 degrees, in good accordance with the values 
presented here.

\section{Discussion and Conclusions}

With the exception of P9, we find the same trends found by
Mukherjee at al. (1999) for PKS~0528+134, i.e.,
that the high states are consistent with a higher bulk
Lorentz factor and lower $\gamma_1$ (low-energy cutoff
of the electron/positron spectrum). However, because the spectra
for 3C~279 contain better simultaneous optical
and X-ray coverage, with detailed X-ray spectral information, 
some of the parameters are better-determined for 3C~279 than for 
PKS~0528+134, in particular the
electron density and the cutoffs $\gamma_1$ and $\gamma_2$.

The broadband spectrum of 3C~279 during P9 constitutes a 
rather special case.
In this spectrum, the level and index of the $\gamma$-ray points
lead to difficulties with the modeling; therefore,
the EGRET data for P9 were examined carefully
to determine whether the effect might be due to unexpected
instrument degradation.  No such unexpected degradation was
found.  The EGRET performance was found to be very similar
to that in P8.  In addition, a spectrum of the Vela pulsar, from 
the EGRET observation immediately prior to P9, was found to be
very similar to those observed previously throughout the CGRO
mission.  Therefore, the P9 EGRET spectrum was assumed to
be correct.

In analogy to our fits to the other epochs, the P9
EGRET spectrum at $\sim 1$~GeV might 
be dominated by Comptonization of direct
accretion disk radiation.  Allowing for a finite
contribution due to Comptonization of reprocessed
accretion disk radiation at $E \gtrsim 10$~GeV, and fitting
simultaneously the optical (synchrotron) and COMPTEL/EGRET
$\gamma$-ray spectra, we find a best-fit electron spectral
index of $p = 2.6$ and a bulk Lorentz factor of $\Gamma = 8$
with the standard accretion disk spectrum used for the fits
to the other observation epochs.  The additional constraints
of the sharp fall-off of the $\gamma$-ray spectrum below
$\sim 10$~MeV and the very hard X-ray spectrum, implying
a $\nu F_{\nu}$ peak energy of the SSC component at
$E_{\rm SSC} \gtrsim 50$~keV, lead to a strong constraint
on $\gamma_1 = 600$ and $n_e \approx 10$~cm$^{-3}$ (for
the fit presented in Fig. 2k, we actually use $n_e
= 9$~cm$^{-3}$).
 
However, with these values for $\gamma_1$ and $\Gamma$
and the standard parameters for the accretion disk luminosity
and the BLR structure, the multi-GeV $\gamma$-ray flux
would be overestimated because of a strongly dominant
contribution from the ECC component in this energy range.
We have tried different combinations of the electron energy
cutoffs, electron density, and the bulk Lorentz factor, but
did not find appropriate parameters to fit the P9 spectrum
with the standard ECC parameters used for the other
fits.  The most convincing way to solve this problem is
to assume variability of the accretion disk.  Specifically,
we assumed that during period P9, 3C~279 had just recovered
from a temporary low-state of the accretion flow.  This
could have lead to a reduced flux in reprocessed accretion
disk emission as the source for the ECC component.  With
a photon energy density in reprocessed accretion disk
radiation of 25\% of the value used for the other observing
periods, we find a satisfactory fit to the P9 broadband
spectrum.

The accretion disk low-state postulated above should have
been quite obvious in an optical spectrum of 3C~279
taken during the P9 EGRET observation, or during the
previous few months.  Attempts have been made to locate
an observer who has taken such a spectrum during or
around the relevant time period.  The only optical spectrum 
known to us (Sillanp\"a\"a et al., private communication) was
taken in June 2000, about four months after the P9
observations.  At that time, 3C~279 showed no lines at all.
Although conditions for that observation were far from ideal,
the result provides
support for the idea that reprocessed accretion disk emission
could have been low at the time of the P9 EGRET observation.

Returning to the ensemble of 11 models, the electron pair 
spectral index $p$ is seen to vary in a way 
that is apparently not related to the $\gamma$-ray flux 
state of the source ($p$ is very well constrained by the
synchrotron spectral index, so this is not an
artifact of a special choice of parameters).
Variations of $n_e$ are of minor importance. In
contrast to the fits to PKS 0528+134, for 3C~279
the high-energy cutoff $\gamma_2$ is rather well
constrained in most of the spectra because of the
simultaneous optical and X-ray coverage and
good X-ray spectral information.

Note that the thermal emission from the 3C~279 accretion disk is 
apparent only during states of low activity.  This has been noted
previously by Pian et al. (1999), who find evidence in IUE data for 
a thermal component underlying the beamed jet only during the 
lowest UV states.

It has been suggested (e.g. Stecker 1999) that 3C~279, being
located relatively nearby (on a cosmological scale) at 
$z = 0.538$, would be a promising
candidate for detection at multi-GeV energies ($\gtrsim 50$~GeV) 
with the new generation of air \v Cerenkov detector facilities, 
such as STACEE, CELESTE, VERITAS, or HESS (for a recent review of 
these detector developments see, e.g., Krennrich 1999), if the 
intrinsic spectrum of 3C~279 is a simple continuation of the 
best-fit power-law to the EGRET spectrum of 3C~279 measured during 
bright $\gamma$-ray flares. However, our model fits predict that 
the intrinsic multi-GeV spectra even in the flaring states of 
3C~279 are rather steep, with photon indices $\gtrsim
3.5$.  Table 4 lists the predicted photon indices and 
integrated fluxes (not accounting for $\gamma\gamma$ absorption 
by the intergalactic infrared background radiation (IIBR)) at 
$E_{\gamma} > 50$~GeV for the 4 $\gamma$-ray high states of 
periods P1, P5a, P5b, and P8. During the lower $\gamma$-ray 
states, no appreciable multi-GeV emission is predicted.

Stecker \& de Jager (1998) calculate a $\gamma\gamma$ absorption 
opacity due to interactions of 50~GeV photons with the IIBR
of $\tau_{\gamma\gamma} \sim$~0.2 -- 0.3, whereas the computations 
of Primack et al. (1999) yield a slightly higher value
of $\tau_{\gamma\gamma} \sim$~0.4 -- 0.5. This leads to predicted $>
50$~GeV fluxes of $\lesssim 10^{-11}$~erg~cm$^{-2}$~s$^{-1}$, even 
during extremely bright $\gamma$-ray flares. This is at least one 
order of magnitude lower than the projected thresholds of STACEE 
and VERITAS, but might be reached by the MAGIC telescope 
(Krennrich 1999).

\acknowledgments

The work of M.~B\"ottcher is supported by NASA through Chandra
Postdoctoral Fellowship Award No. 9-10007, issued by
the Chandra X-ray Center, which is operated by the
Smithsonian Astrophysical Observatory for and on behalf
of NASA under contract NAS 8-39073.

The Nordic Optical Telescope is operated on the island of La Palma 
jointly by Denmark, Finland, Iceland, Norway and Sweden in the 
Spanish Observatorio del Roque de
Los Muchachos of the Instituto de Astrofisica de Canarias.
The Tuorla Observatory authors wish to thank The Academy of Finland for
support. 

The work at Torino Observatory was partly supported by the Italian
Ministry for University and Research (MURST) under grant Cofin98-02-32 
and by the Italian Space Agency (ASI).

The Georgia State University authors wish to thank Lowell and
Mount Stromlo / Siding Spring Observatories for allocations of 
observing time.  This work has been supported in part by an award 
from GSU's RPE Fund to PEGA, and by grants from the Research 
Corporation and NASA (NAGW-4397).  

D. Backman was a visiting astronomer at
CTIO/NOAO, operated by AURA under a cooperative agreement with
the NSF.  The National Undergraduate Research Observatory
(NURO) is operated by Lowell Observatory under an agreement
with Northern Arizona University and the NURO consortium.
D. Backman and Franklin \& Marshall College thank the University
of Delaware Space Grant Colleges consortium for funding in support of 
NURO membership and observing runs.

For this work, P. Jogee was supported by NSF grant AST 99-81546
to OVRO.

The quasar monitoring at the Mets\"ahovi Radio Observatory has been partly
financed by The Academy of Finland.

H. Bock, J. Heidt and S.J. Wagner acknowledge support by the DFG 
(SFB 328 and 439), and CAHA/DSAZ
for support during several observing runs on Calar Alto.

O.M. Kurtanidze thanks the Astrophysikalisches Institute Potsdam for 
support.

The JCMT is operated by the Joint Astronomy Centre in Hilo, Hawaii on
behalf of the parent organizations Particle Physics and Astronomy Research
Council in the United Kingdom, the National Research Council of Canada and
The Netherlands Organization for Scientific Research. 

The operation and data analysis for the COMPTEL, EGRET, and
OSSE instruments on the Compton Gamma Ray Observatory were 
supported by NASA.

IRAF is distributed by the National
Optical Astronomy Observatories, which is operated by the Association
of Universities for Research in Astronomy, Inc., under cooperative
agreement with the National Science Foundation.

\clearpage

\begin{table}
\begin{center}
Table 1 - Observation intervals \\
\begin{tabular}{lll}
\\ \hline\hline \\[-.11in]
Interval & Start date & End date \\ \hline \\[-.11in]
P1    & 1991 Jun 15 & 1991 Jun 28 \nl
P2    & 1992 Dec 22 & 1993 Jan 12 \nl
P3a   & 1993 Oct 19 & 1993 Dec 01 \nl
P3b   & 1993 Dec 13 & 1994 Jan 03 \nl
P4    & 1994 Nov 29 & 1995 Jan 10 \nl
P5a   & 1996 Jan 16 & 1996 Jan 30 \nl
P5b   & 1996 Jan 30 & 1996 Feb 06 \nl
P6a   & 1996 Dec 10 & 1997 Jan 28 \nl
P6b   & 1997 Jun 17 & 1997 Jun 24 \nl
P8    & 1999 Jan 19 & 1999 Feb 01 \nl
P9    & 2000 Feb 08 & 2000 Mar 01 \nl
\hline
\end{tabular}
\end{center}
\end{table}                                         

\begin{table}
\begin{center}
Table 2 - Model parameters fixed for all epochs \\
\begin{tabular}{ll}
\\ \hline\hline \\[-.11in]
Parameter & Value \\ \hline \\[-.11in]
Accretion disk luminosity\tablenotemark{\dag} & $10^{46}$ erg s$^{-1}$ \nl
Observing angle wrt jet   & 2$^\circ$ \nl
Thomsom depth of BLR      & 0.003 \nl
Inner boundary of BLR     & 0.1 pc \nl
Outer boundary of BLR     & 0.4 pc \nl
Blob injection height     & 0.025 pc \nl
Blob radius               & 6$\times$10$^{16}$ cm \nl
Magnetic field            & 1.5 G \nl

\end{tabular}

\tablenotetext{\dag}{Although the parameters above were used also
for P9, in that case only, it was necessary to reduce artificially
the density of reprocessed disk photons to 25\% of that calculated
by the computer code.}
\end{center}
\end{table} 

\begin{deluxetable}{clrrccc}
\tablenum{3}
\tablecaption{Summary of parameters from fits}
\tablewidth{480pt}
\tablehead{
\colhead{Observation} & \colhead{$\gamma$ ray state} & \colhead{$\gamma_1$}   
  & \colhead{$\gamma_2$} & \colhead{$p$}  & \colhead{$n_e$} 
  & \colhead{$\Gamma_{jet}$}
}
\startdata
P1 & high; moderate flare   & 350 & 100$\times$10$^{3}$ & 1.90 & 28 & 14 \nl
P2 & very low               &1200 &  14$\times$10$^{3}$ & 2.95 & 30 &  6 \nl
P3a& moderate               & 700 &  15$\times$10$^{3}$ & 2.10 & 25 &  6 \nl
P3b& moderate               &1500 &  20$\times$10$^{3}$ & 3.10 & 15 & 5.5\nl
P4 & low                    &1500 &  50$\times$10$^{3}$ & 3.10 & 40 & 10 \nl
P5a& high; some variability & 600 & 100$\times$10$^{3}$ & 2.40 & 20 & 10 \nl
P5b& very large flare       & 400 & 100$\times$10$^{3}$ & 3.00 & 35 & 13 \nl
P6a& low                    & 750 &  30$\times$10$^{3}$ & 2.50 & 10 &  4 \nl
P6b& moderate               &1350 &  15$\times$10$^{3}$ & 2.70 & 10 & 5.5\nl
P8 & high; some variability & 450 & 100$\times$10$^{3}$ & 2.20 & 10 &  6 \nl
P9 & high; some variability & 600 &  50$\times$10$^{3}$ & 2.60 &  9 &  8 \nl
							 
\enddata 
							  
\end{deluxetable}

\begin{table}
\begin{center}
Table 4.\ \ Predicted photon spectral indices $\alpha$ and integrated
	unabsorbed $> 50$~GeV fluxes from our model fits to the 
	$\gamma$-ray high states of 3C~279 \\
\begin{tabular}{ccc}
\\ \hline\hline \\[-.11in]
Period & $\alpha$ & $F_{> 50 \, {\rm GeV}}$ [erg~cm$^{-2}$~s$^{-1}$] \\
\hline
\\[-.11in]
P1  & 5.0 & $1.2 \times 10^{-11}$   \nl
P5a & 5.0 & $1.3 \times 10^{-12}$   \nl
P5b & 3.5 & $3.3 \times 10^{-12}$   \nl
P8  & 5.0 & $6.7 \times 10^{-13}$   \nl
\hline
\end{tabular}
\end{center}
\end{table}

\clearpage

\figcaption{3C~279 light curve in $\gamma$ rays $>$100 MeV, 1991 to 
2000.  Horizontal error bars show the length of each observation. 
\label{fig1}}

\figcaption{Broadband 3C~279 spectra for the eleven time intervals listed 
in Table 1.  Filled triangle with error bars, single measurement (or 
upper limit); vertical bar with no triangle, range of two or more 
measurements, including errors.  The curves are for the model described
in section 4 and Tables 2 and 3: short-dashed line ($\sim 10^{15}$ Hz), 
thermal emission from the
accretion disk; short-dashed line ($\sim 10^{21}$ Hz), disk emission
comptonized in the jet; long-dashed line, synchrotron radiation from the
jet; dotted line, synchrotron self-Compton radiation from the jet;
short-long-dashed line, BLR emission comptonized in the jet.
\label{fig2} }

\figcaption{Schematic showing the geometric properties of the model.
\label{fig3} }

\clearpage
\begin{figure}
\figurenum{1}
\epsscale{1.5}
\plotfiddle{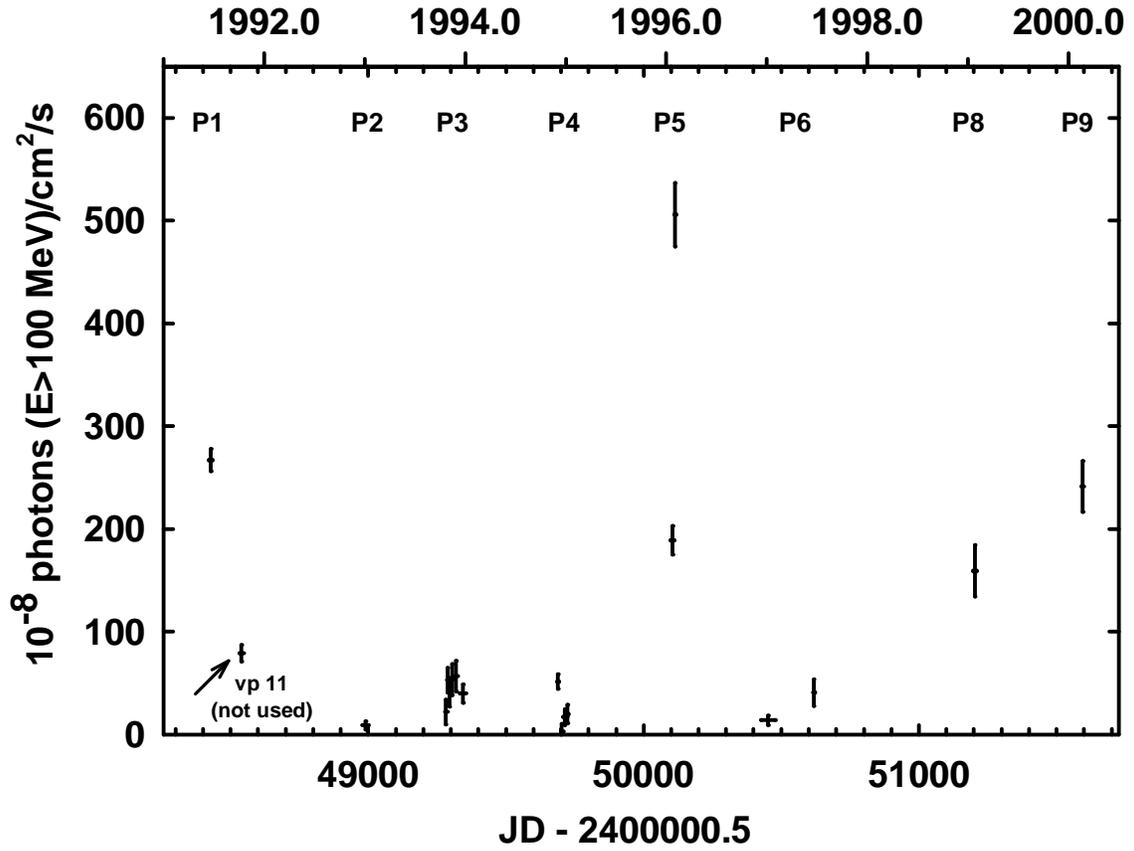}{200pt}{0}{100}{100}{-365pt}{-150pt}
\caption{3C~279 light curve in $\gamma$ rays $>$100 MeV, 1991 to
2000.  Horizontal error bars show the length of each observation.}
\end{figure}

\clearpage
\begin{figure}
\figurenum{2a}
\epsscale{1.0}
\plotfiddle{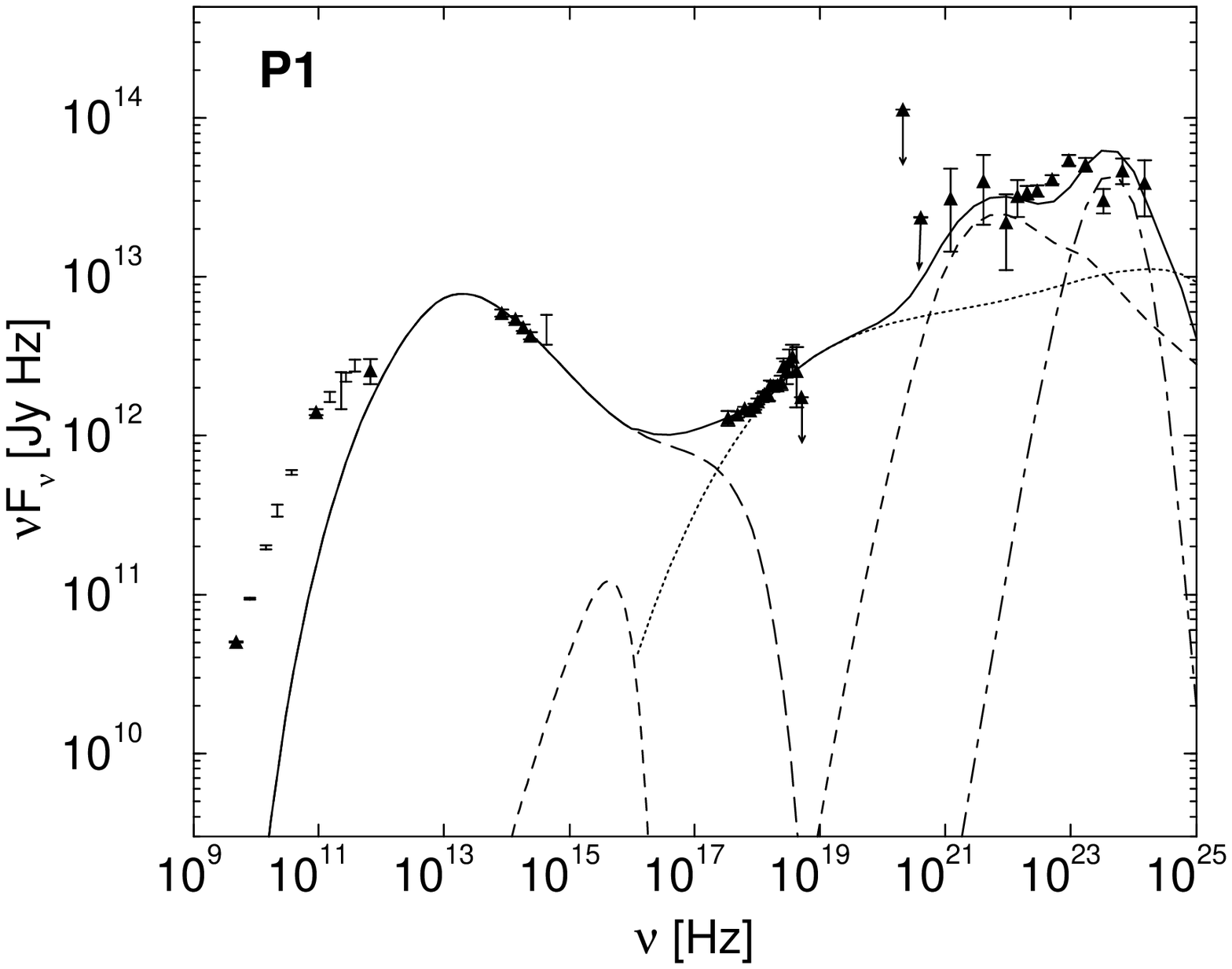}{200pt}{0}{90}{90}{-250pt}{-30pt}
\caption{ }
\end{figure}

\clearpage 
\begin{figure} 
\figurenum{2b} 
\epsscale{1.0} 
\plotfiddle{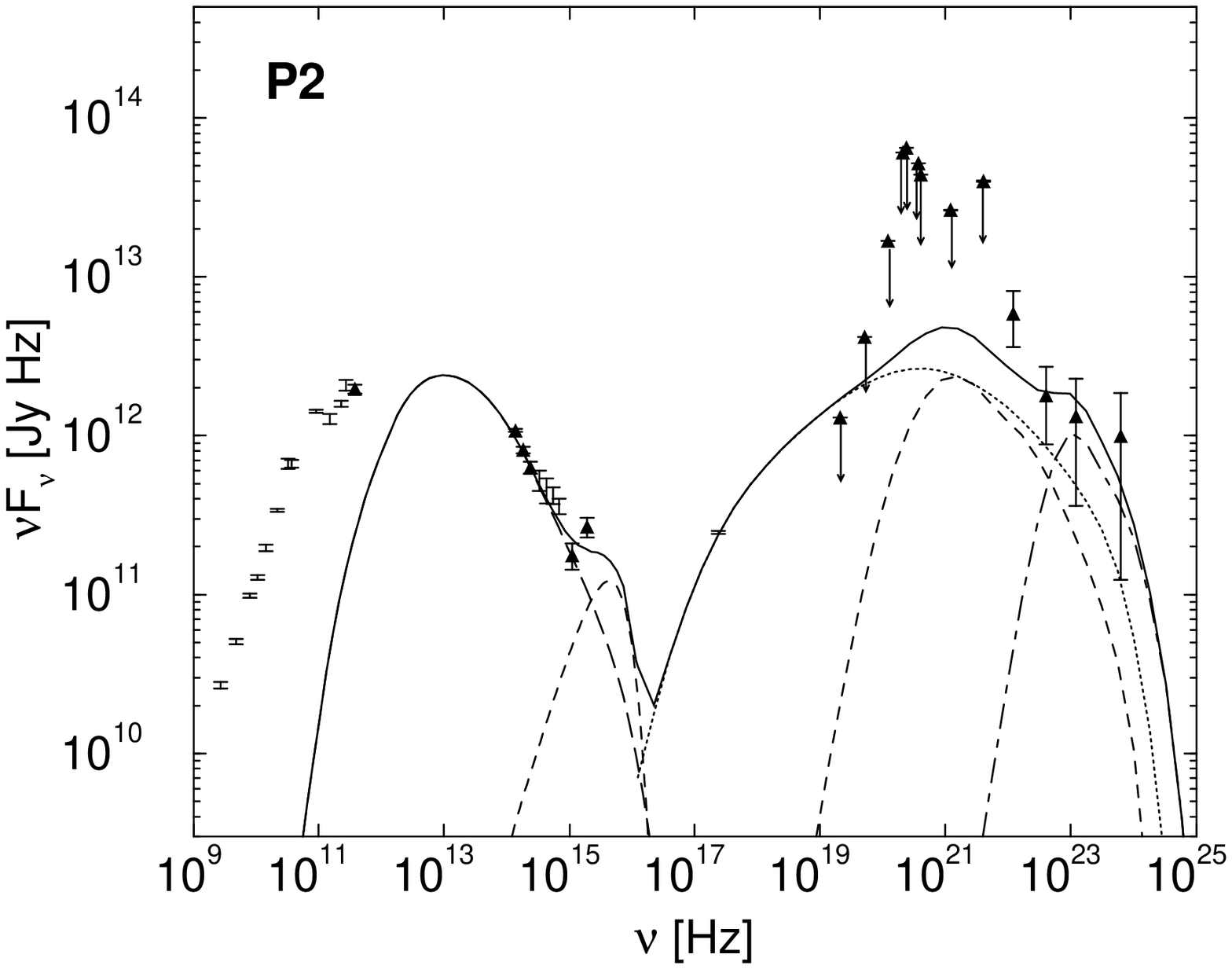}{200pt}{0}{90}{90}{-250pt}{-30pt} 
\caption{ }
\end{figure}

\clearpage  
\begin{figure}  
\figurenum{2c} 
\epsscale{1.0}  
\plotfiddle{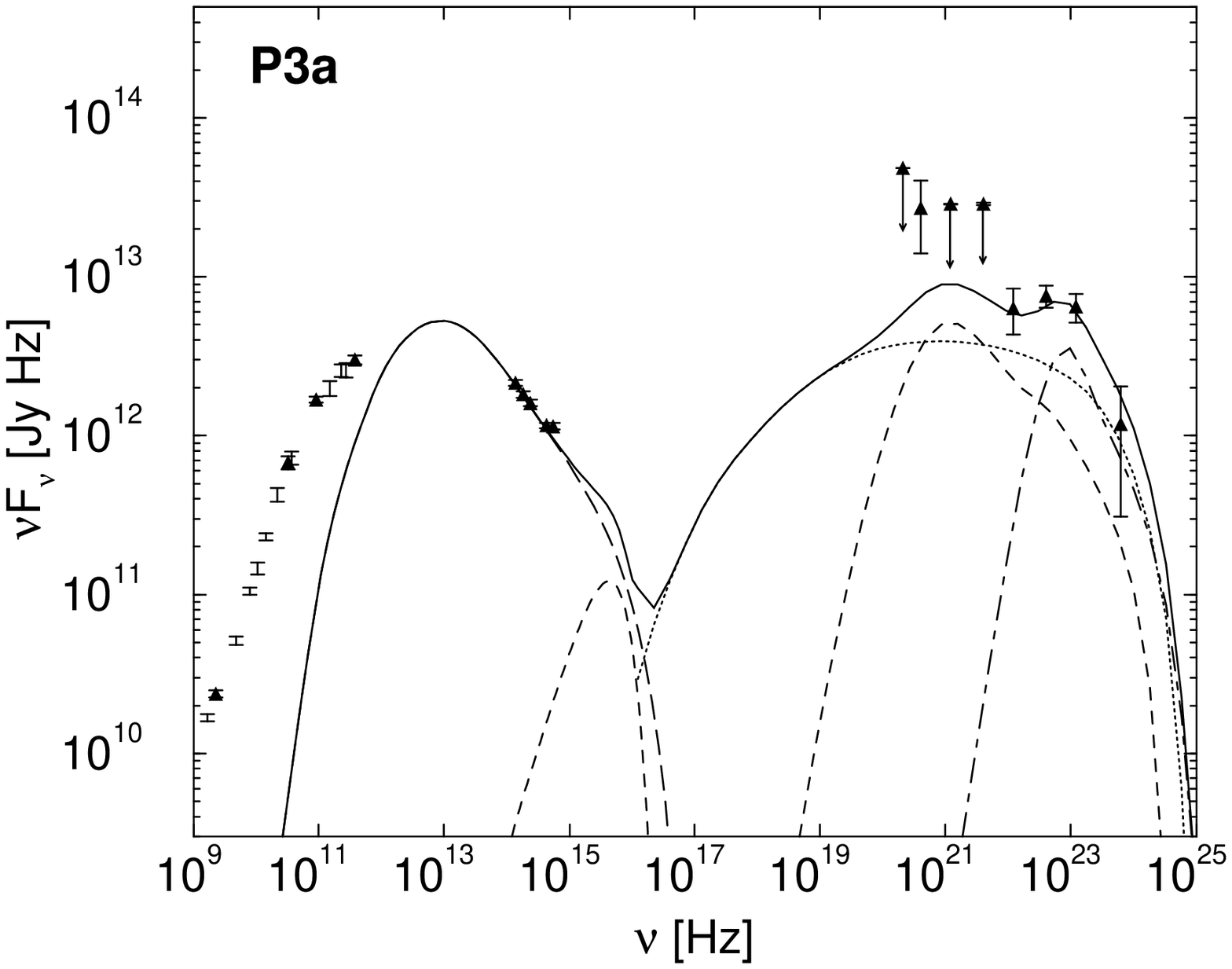}{200pt}{0}{90}{90}{-250pt}{-30pt}  
\caption{ } 
\end{figure}

\clearpage  
\begin{figure}  
\figurenum{2d} 
\epsscale{1.0}  
\plotfiddle{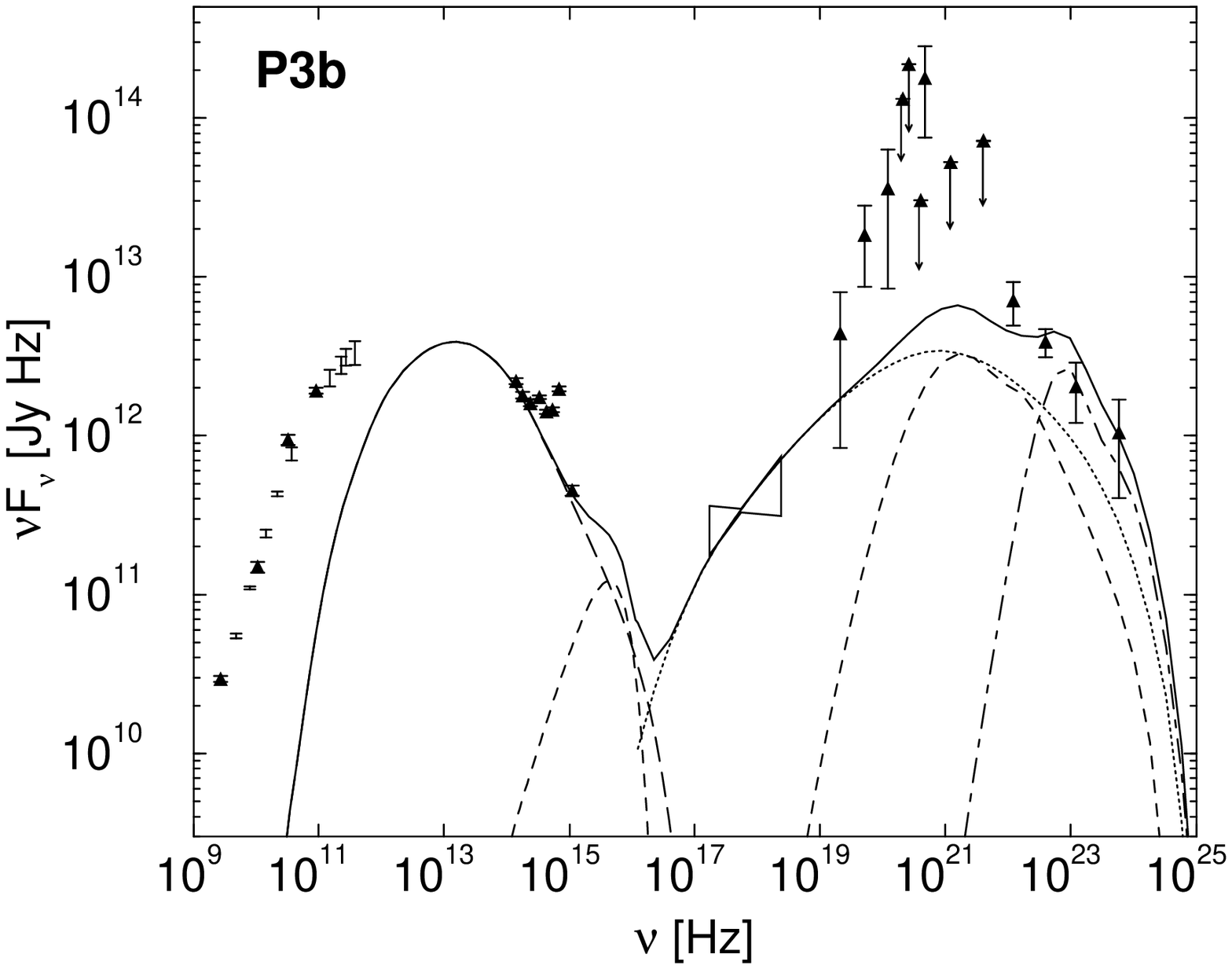}{200pt}{0}{90}{90}{-250pt}{-30pt}  
\caption{ } 
\end{figure}

\clearpage  
\begin{figure}  
\figurenum{2e} 
\epsscale{1.0}  
\plotfiddle{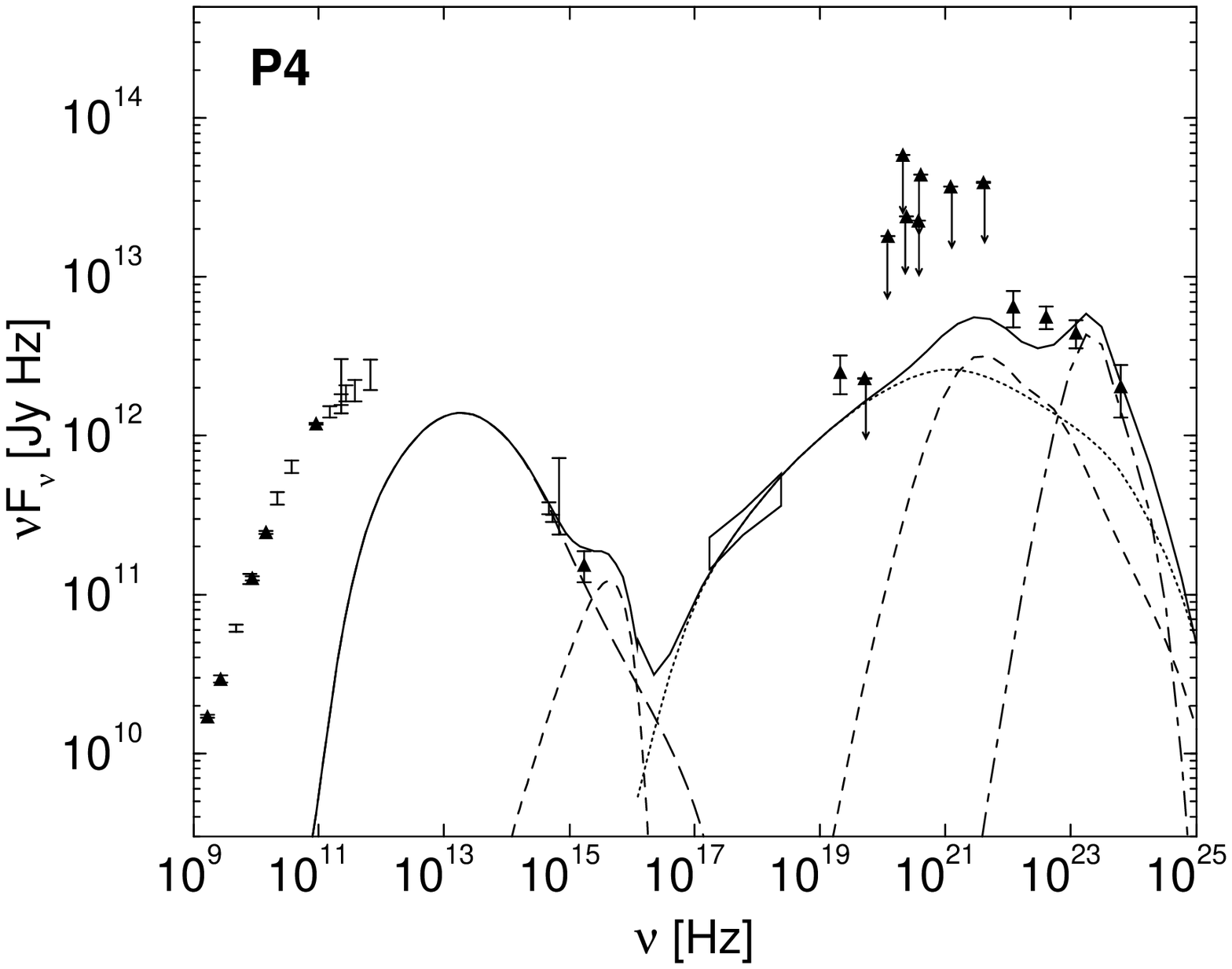}{200pt}{0}{90}{90}{-250pt}{-30pt}  
\caption{ } 
\end{figure}

\clearpage  
\begin{figure}  
\figurenum{2f} 
\epsscale{1.0}  
\plotfiddle{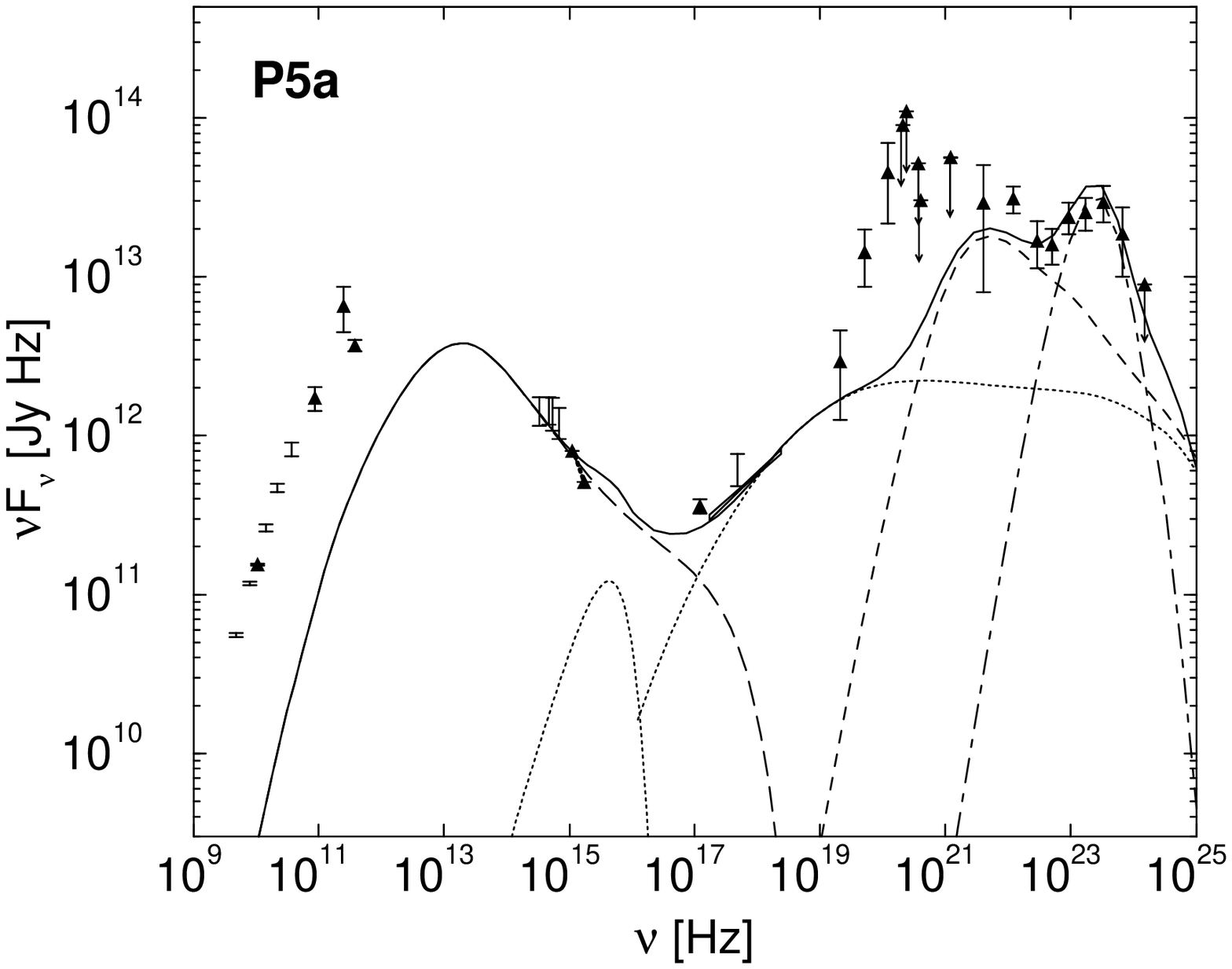}{200pt}{0}{90}{90}{-250pt}{-30pt}  
\caption{ } 
\end{figure}

\clearpage  
\begin{figure}  
\figurenum{2g} 
\epsscale{1.0}  
\plotfiddle{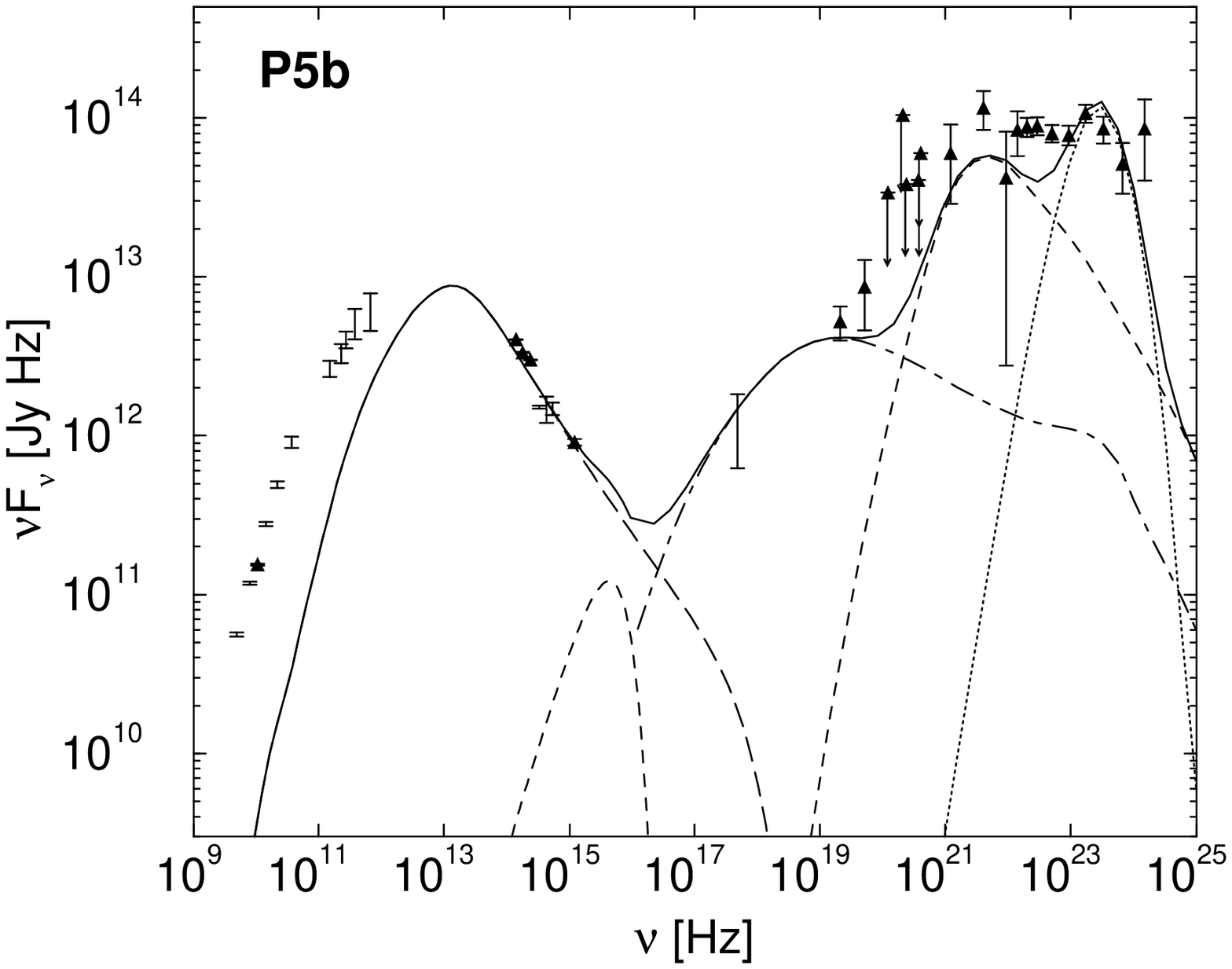}{200pt}{0}{90}{90}{-250pt}{-30pt}  
\caption{ } 
\end{figure}

\clearpage  
\begin{figure}  
\figurenum{2h} 
\epsscale{1.0}  
\plotfiddle{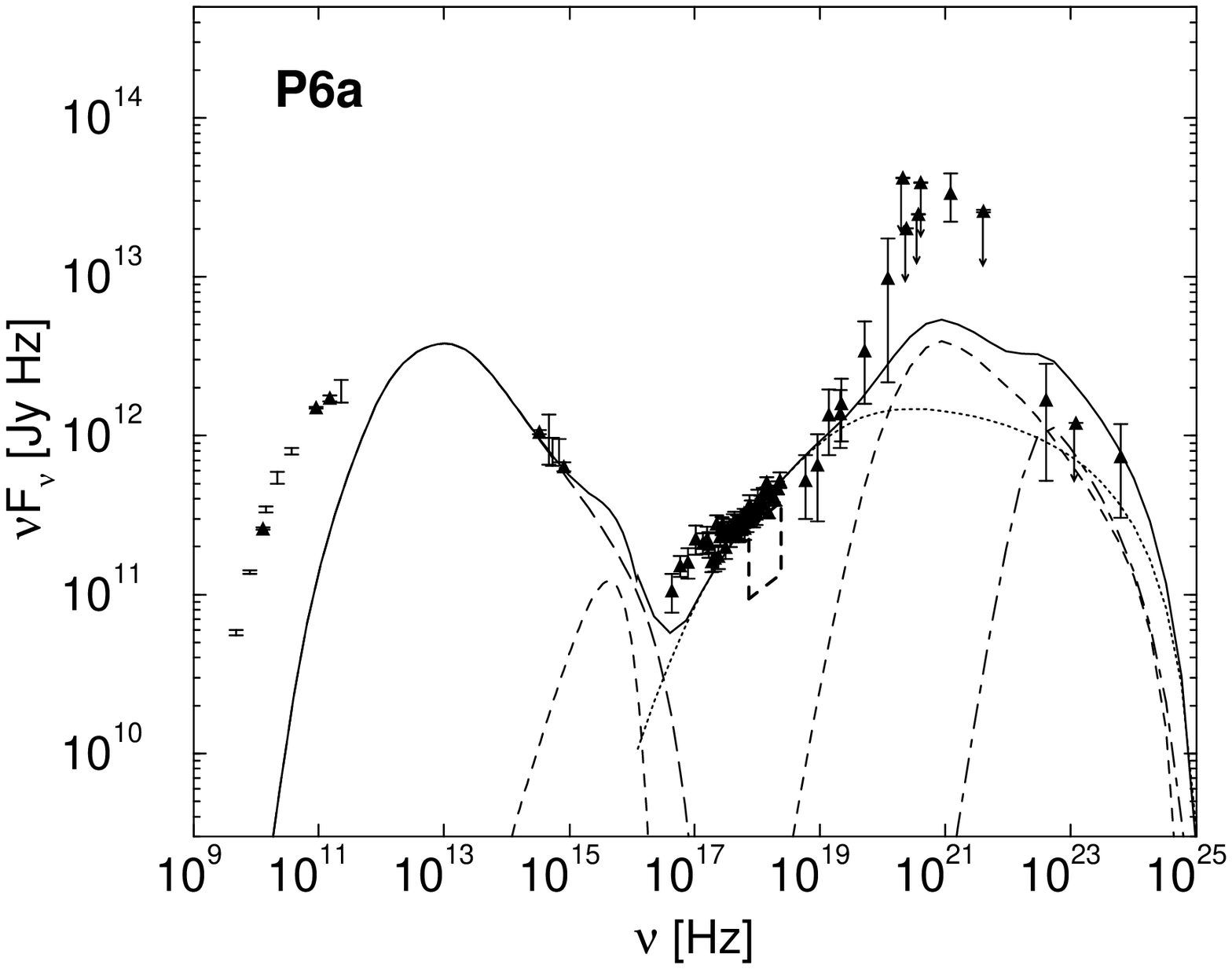}{200pt}{0}{90}{90}{-250pt}{-30pt}  
\caption{ } 
\end{figure}

\clearpage  
\begin{figure}  
\figurenum{2i} 
\epsscale{1.0}  
\plotfiddle{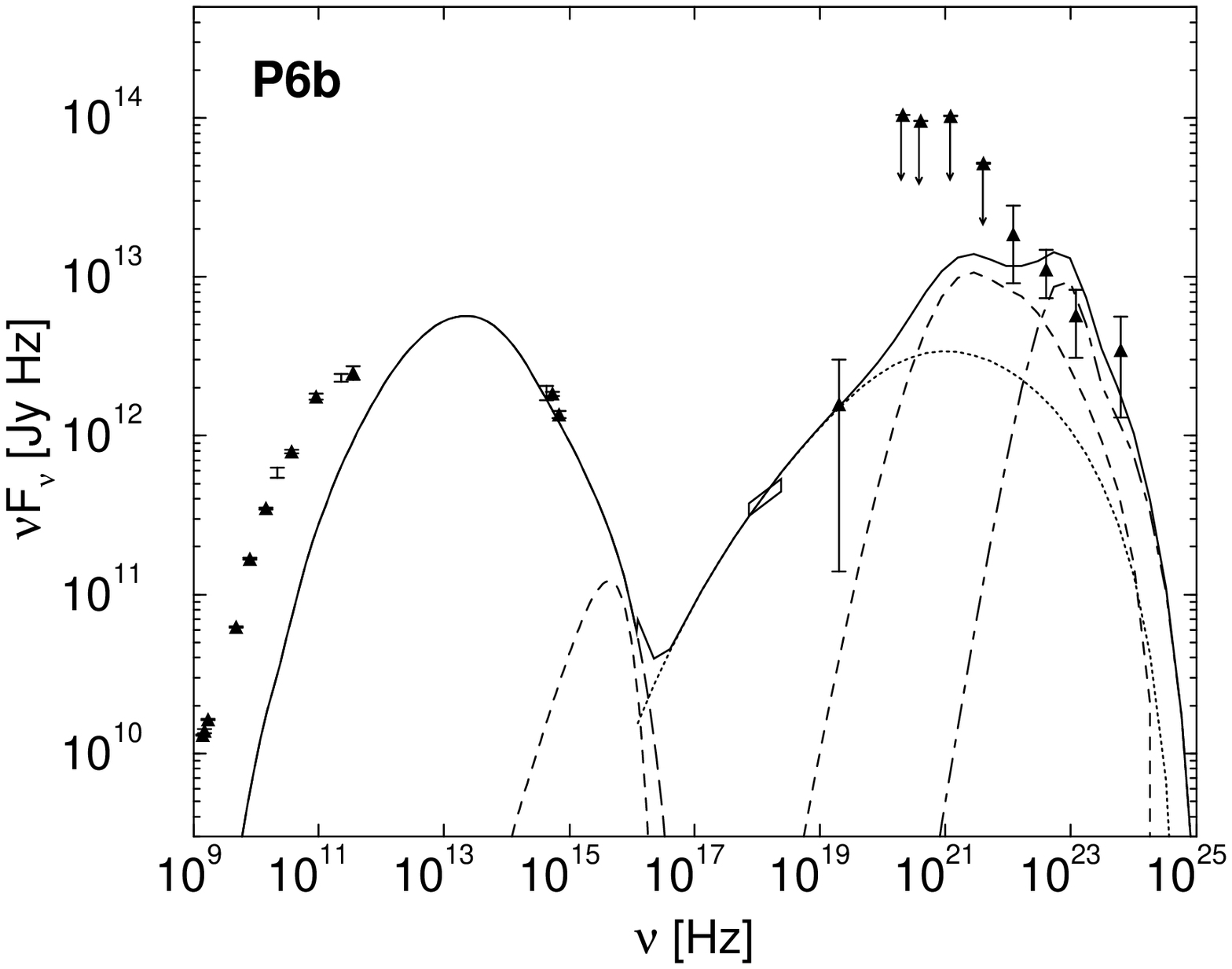}{200pt}{0}{90}{90}{-250pt}{-30pt}  
\caption{ } 
\end{figure}

\clearpage  
\begin{figure}  
\figurenum{2j} 
\epsscale{1.0}  
\plotfiddle{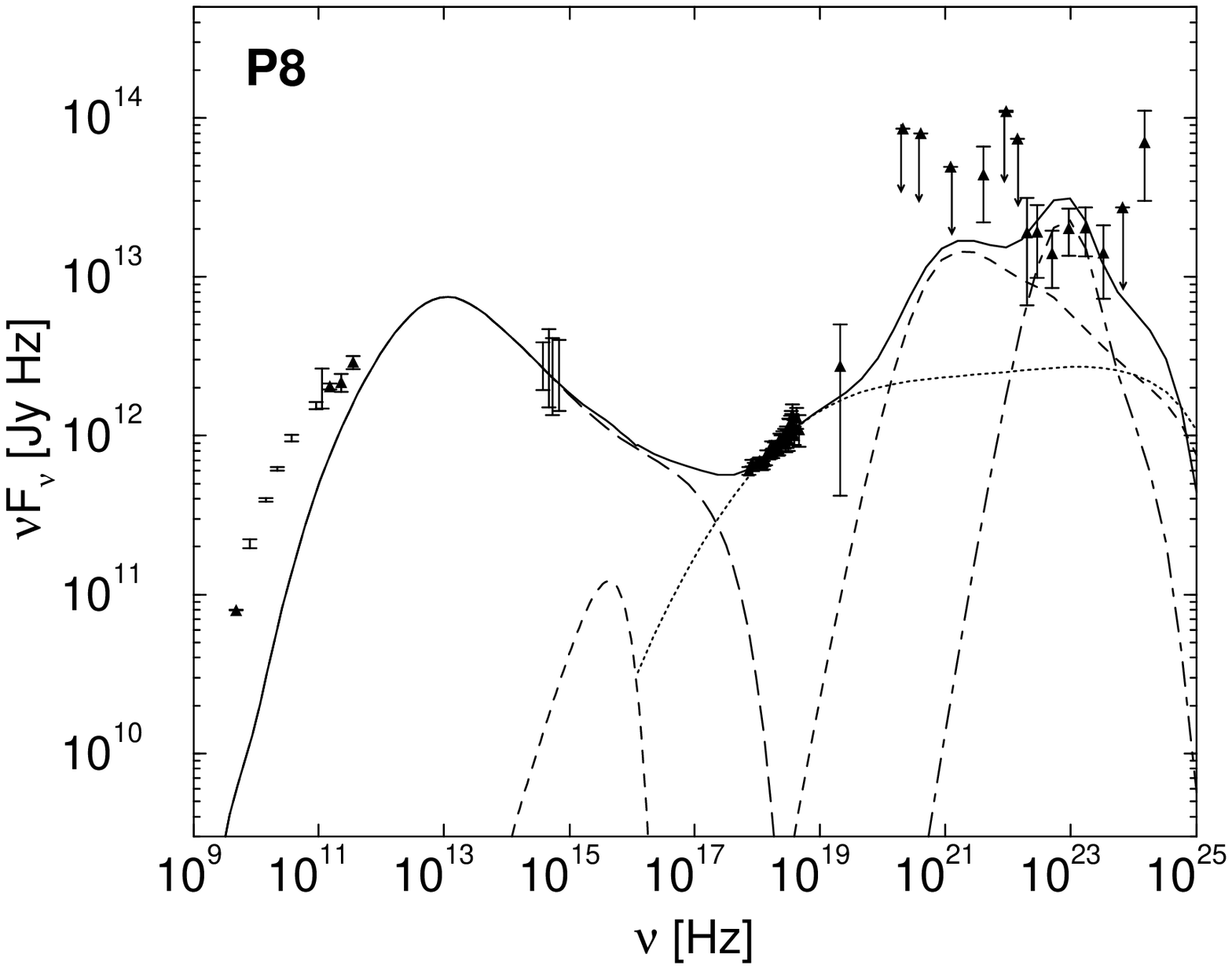}{200pt}{0}{90}{90}{-250pt}{-30pt}  
\caption{ } 
\end{figure}

\clearpage  
\begin{figure}  
\figurenum{2k} 
\epsscale{1.0}  
\plotfiddle{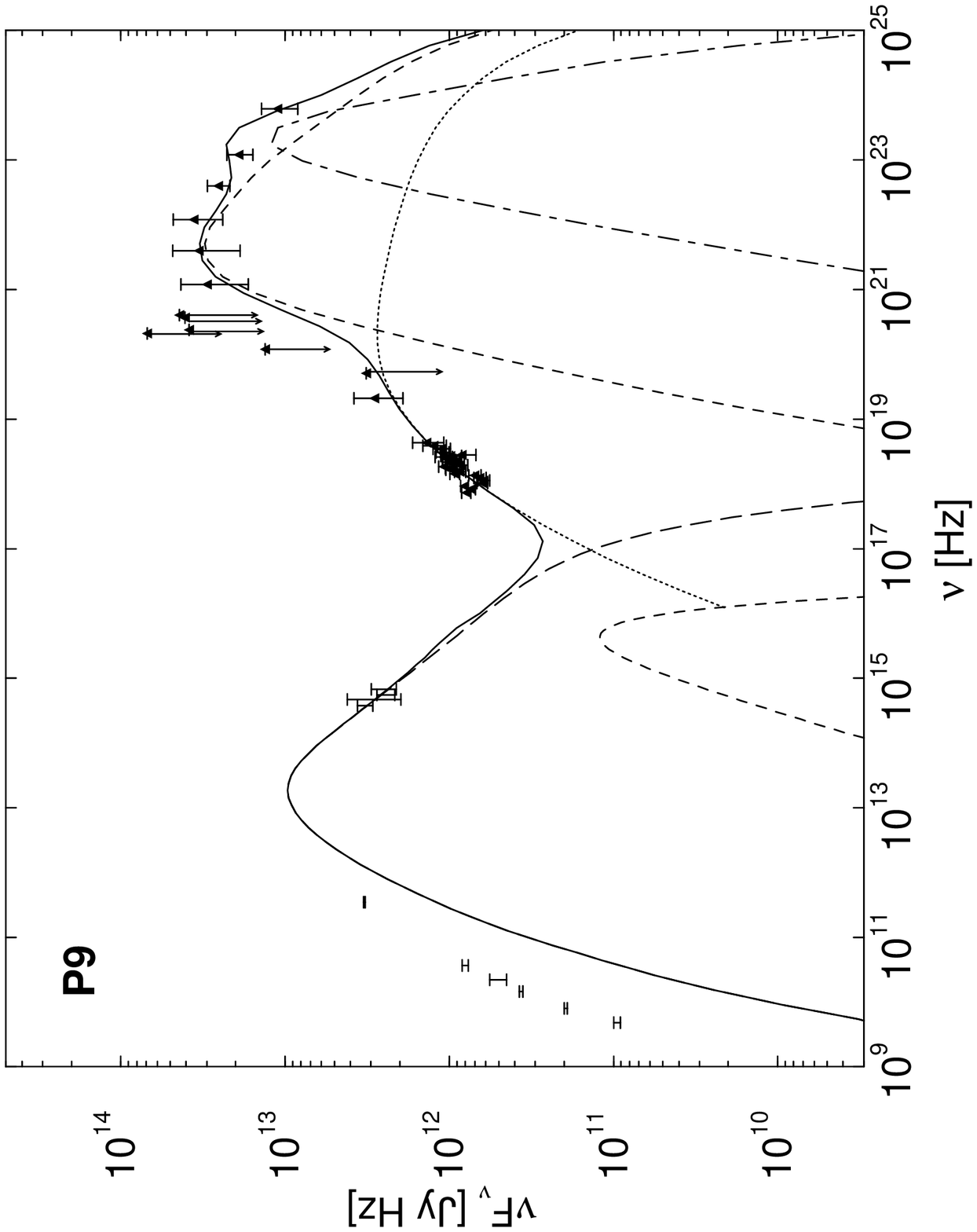}{200pt}{-90}{77}{77}{-250pt}{430pt}  
\caption{ } 
\end{figure}

\clearpage
\begin{figure}
\figurenum{3}
\epsscale{1.0}
\plotfiddle{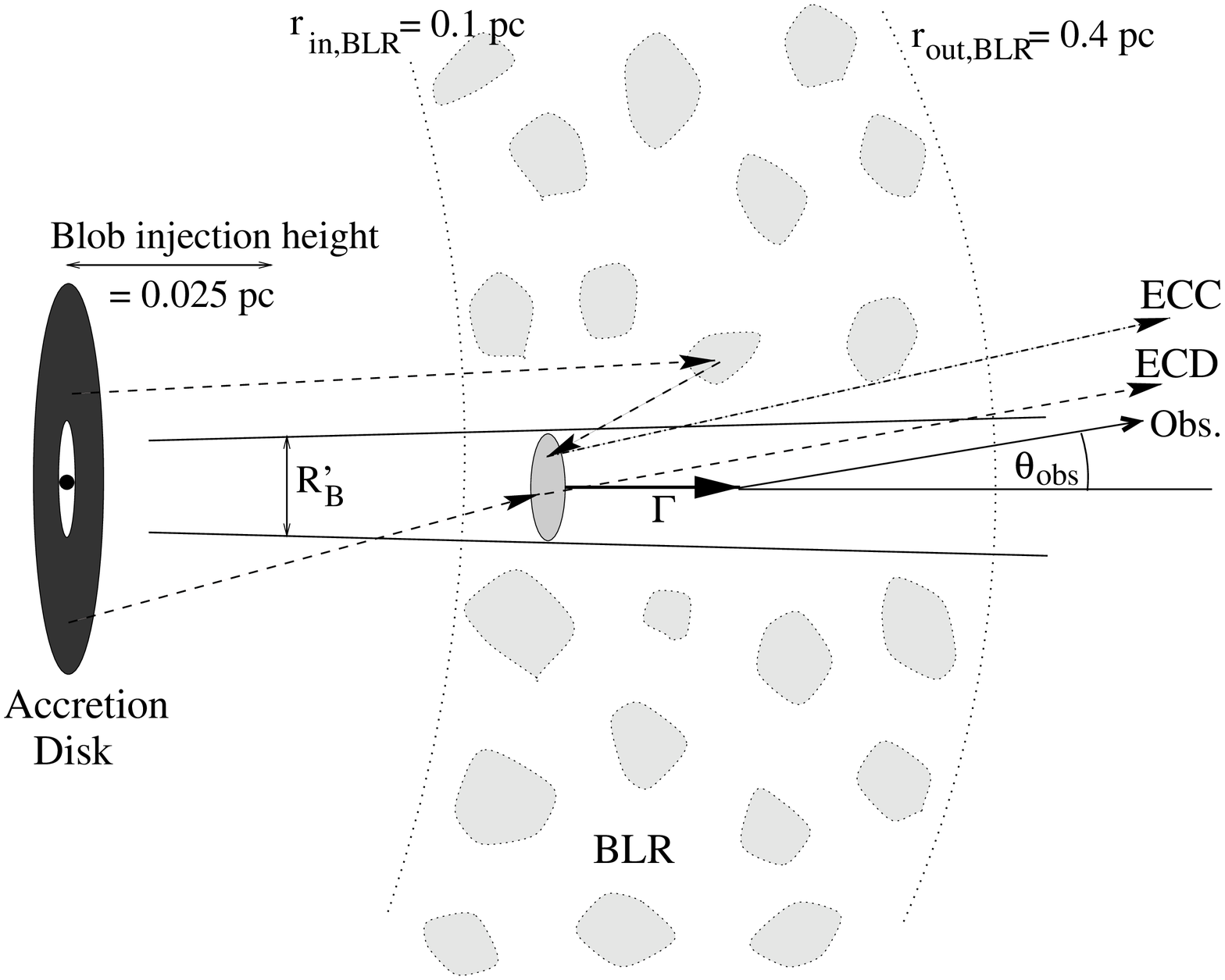}{200pt}{0}{55}{55}{-195pt}{20pt}
\caption{Schematic showing the geometric properties of the model.}
\end{figure}

\end{document}